\documentclass[11pt, a4paper, twoside]{article}
\usepackage[a4paper]{anysize}

\usepackage{amssymb}
\usepackage{booktabs}
\usepackage{amsmath}
\usepackage{amsfonts, amsthm, graphicx, stmaryrd, fancyhdr}

\usepackage{enumerate}
\usepackage{color}

\newcommand{\om}{\omega}
\newcommand{\bmo}{\textrm{BMO}}
\newcommand{\al}{\alpha}

\newcommand{\eps}{\varepsilon}
\renewcommand{\theta}{\vartheta}
\newcommand{\lam}{\lambda}

\newcommand{\dist}{\mathrm{dist}}

\newcommand{\IN}{\mathbb{N}}

\newcommand{\IR}{\mathbb{R}}
\newcommand{\R}{\mathbb{R}}

\newcommand{\cA}{\mathcal{A}}
\newcommand{\cB}{\mathcal{B}}

\newcommand{\cD}{\mathcal{D}}
\newcommand{\cE}{\mathcal{E}}
\newcommand{\cF}{\mathcal{F}}

\newcommand{\cH}{\mathcal{H}}

\newcommand{\cL}{\mathcal{L}}

\newcommand{\cP}{\mathcal{P}}
\newcommand{\cQ}{\mathcal{Q}}

\newcommand{\cS}{\mathcal{S}}

\newcommand{\be}{\begin{eqnarray*}}
\newcommand{\ee}{\end{eqnarray*}}
\newcommand{\ben}{\begin{eqnarray}}
\newcommand{\een}{\end{eqnarray}}

\theoremstyle{plain}
\newtheorem{theo}{Theorem}[section]
\newtheorem{lemma}[theo]{Lemma}

\newtheorem{corollary}[theo]{Corollary}

\theoremstyle{definition}
\newtheorem{defi}[theo]{Definition}

\newtheorem{remark}[theo]{Remark}
\newtheorem{example}[theo]{Example}

\title{Pricing and hedging of derivatives based on non-tradable underlyings}
\author{Stefan Ankirchner and Peter Imkeller and Gon{\c c}alo Dos Reis
\\ Institut f\"ur Mathematik\\ Humboldt-Universit\"at zu Berlin\\
Unter den Linden 6\\ 10099 Berlin\\ Germany }
\begin{document}

\maketitle
\begin{abstract}
This paper is concerned with the study of insurance related
derivatives on financial markets that are based on non-tradable
underlyings, but are correlated with tradable assets. We calculate
exponential utility-based indifference prices, and corresponding
derivative hedges. We use the fact that they can be represented in
terms of solutions of forward-backward stochastic differential
equations (FBSDE) with quadratic growth generators. We derive the
Markov property of such FBSDE and generalize results on the
differentiability relative to the initial value of their forward
components. In this case the optimal hedge can be represented by
the price gradient multiplied with the correlation coefficient.
This way we obtain a generalization of the classical 'delta hedge'
in complete markets.
\end{abstract}



\section*{Introduction}
In recent years more and more financial instruments have been
created which are not derived from exchange traded securities. For
instance in 1999 the Chicago Mercantile Exchange introduced
weather futures contracts, the payoffs of which are based on
average temperatures at specified locations. Another example of
derivatives with non-tradable underlyings are catastrophe futures
based on an insurance loss index regulated by an independent
agency or simply derivatives based on equity indices such as S\&P
or DAX.\par\smallskip

Financial or insurance derivatives of this type are impossible to
perfectly hedge, since it is impossible to trade the underlying
variable that carries independent uncertainty. To circumvent this
problem, in practice one looks for a tradable asset that is
correlated to the non-tradable underlying of the derivative. Even
though investing in the correlated asset cannot provide a total
hedge of the derivative, and a non-hedgeable \emph{basis risk}
remains, it is better than not hedging at all.\par\smallskip

In the following we will investigate utility-based pricing
principles for derivatives based on non-tradable underlyings.
Moreover we will show how the derivatives can be partially hedged
by investing in correlated assets. We present {\em explicit}
hedging strategies that optimize the expected utility of a
portfolio of such derivatives. To this end we will establish some
structure and smoothness properties of indifference prices such as
the Markov property and differentiability with respect to the
underlyings. Once these properties are established, we can
explicitly describe the optimal hedging strategies in terms of the
price gradient and correlation coefficients. This way we obtain a
generalization of the classical {\em delta hedge} of the
Black-Scholes model.\par\smallskip

The hedging of claims based on non-tradable underlyings has
already been studied by many authors, see for example
\cite{henhob02}, \cite{henderson}, \cite{muszar}, \cite{davis2},
\cite{monoyios}, \cite{05AIP}. As a common feature of all these
papers, optimal hedging strategies are derived with standard
stochastic control techniques. The essential components of this
analytical approach consist in a formulation of the optimization
problem in terms of HJB partial differential equations, and the
use of a verification theorem and uniqueness result in order to
obtain a representation of the indifference price and the optimal
control strategy. We instead employ an approach with a stochastic
focus. It starts with the well-known observation that the maximal
expected \emph{exponential} utility may be computed by appealing
to the martingale optimality principle which leads to a
description of price and optimal hedging strategy in terms of a
{\em forward-backward stochastic differential equation} (FBSDE)
with a nonlinearity of quadratic type (see \cite{rouge},
\cite{huimmu}). This immediately implies that the utility
indifference price resp. hedge is equal to the difference of
initial states resp. control processes of two FBSDE with a
quadratic nonlinearity in the generator. The forward component is
given by a Markov process describing the non-tradable underlying.
The main mathematical contribution of this paper is that it
provides simple sufficient conditions for general FBSDE with
quadratic nonlinearity to satisfy a Markov property, and - for the
BSDE component - to be differentiable with respect to the initial
condition of the forward equation. The techniques for proving
differentiability of BSDE with quadratic nonlinearity have been
developed independently in \cite{bricon} and \cite{animre}.
Unfortunately, the setup of both papers is not general enough to
cover the BSDE needed to calculate exponential indifference
prices. Therefore, a slight generalization of these
differentiability results is given in the last section of this
paper.\par\smallskip

As a consequence of the explicit description of indifference
prices and hedges in terms of the solution processes of the FBSDE,
and in view of the smoothness results mentioned, it is
straightforward to describe optimal hedging strategies in terms of
the indifference price gradient and the correlation coefficients
explicitly. An economics related contribution of the paper is that
the framework presented allows to refine the results obtained for
example in \cite{muszar}, \cite{davis2}. Firstly, no longer we
need to impose any restrictions on the coefficients of the
diffusion modeling the tradable asset price. More importantly, the
BSDE techniques allow to deal with multidimensional underlyings
and traded assets. In the approach based on the HJB equation, a
solution of the PDE is obtained by using an exponential Hopf-Cole
transformation that in general seems to require that there exists
only one traded asset. In practice many derivatives are based on
more than one underlying, such as spread options or basket
options. In order to illustrate how to hedge with more than one
asset, we will study in more detail so-called crack spreads, which
are written for instance on the difference of crude oil futures
and kerosene prices (see Example \ref{ex.kerosene} and
\ref{ex.ker2}).\par\smallskip

Finally we address the pricing of derivatives by the
\emph{marginal utility approach}. If a company wishes to trade
risk not covered by securities on an exchange, they are forced to
go outside the exchange to get tailored products to serve their
specific needs. These deals that do not go through the exchange
trading (although the underlyings may be traded there) and are
done directly between buyer and seller are called
\emph{over-the-counter} (OTC). For example, airlines regularly
make this kind of OTC deals in order to protect themselves against
kerosene price fluctuations, which underlines that the amount of
money involved in this type of deals is non-negligible! Investment
banks offering OTC deals face the problem of finding a fair price
of these agreements. Indifference prices are often a reasonable
solution. However, they are not linear! The standard way out, as
suggested in the economics literature, is pricing by marginal
utility. The marginal utility price is the differential quotient
of the indifference price with respect to a marginal amount of the
derivative. Here again the first thing to verify is the
differentiability of the FBSDE. This in turn allows to derive the
dynamics of the marginal utility price as a BSDE with a driver
satisfying a random Lipschitz condition.\par\smallskip

BSDE with generators of quadratic nonlinearity in the control
variable (which will in the sequel sometimes simply be called
\emph{quadratic BSDE}) are described by equations of the type
$$Y_t = \xi + \int_t^T f(s,Y_s,Z_s) ds - \int_t^T Z_s d W_s,\quad
0\le t\le T,$$ where $f$ is a predictable function satisfying
$|f(t,y,z)|\leq C(1+|y|+|z|^2)$ with some constant $C$. Our
differentiability results are based on the assumption that the
derivative to be hedged, denoted by $\xi$, is essentially bounded.
This guarantees that the integral process $\int_0^\cdot Z dW$ is a
so-called BMO martingale, and hence the density process of a new
equivalent probability measure, say $Q$. By switching to the
measure $Q$ one can derive moment estimates needed in order to
prove differentiability. The assumption that the derivative has to
be bounded seems to be a disadvantage of using BSDE in the
stochastic approach instead of working with the HJB partial
differential equation in the analytical approach. In practice,
this is of no importance.\par\smallskip


The paper is organized as follows: in Section \ref{model} we
introduce the model, in Section \ref{controlviabsdes} we briefly
recall results from \cite{huimmu} concerning the solution of the
problem of exponential expected utility maximization in terms of
stochastic control problems and FBSDE with nonlinearities of
quadratic type. In Section \ref{markovofprice} we show structure
properties of indifference prices of derivatives based on a
non-tradable Markovian index process. In Section
\ref{explicithedge} we derive explicit formulas for the optimal
hedges of such derivatives, and in Section \ref{section:MUP} we
describe the dynamics of the marginal utility price. All the
economics related results are based on mathematical properties of
quadratic FBSDE, which will be proved in the last section.

\section{The model}\label{model}
Let $d\in \IN$ and let $W$ be a $d$-dimensional Brownian motion on a probability space $(\Omega, \cF, P)$. We
denote by $(\cF_t)$ the completion of the filtration generated by $W$. Suppose that a derivative with maturity
$T>0$ is based on a $\R^m$-dimensional non-tradable index (think of a stock, temperature or loss index) with
dynamics
\begin{equation}\label{riskpro}
dR_t = b(t, R_t)dt + \rho(t, R_t) dW_t,
\end{equation}
where $b: [0,T] \times \R^m \to \R^m$ and $\rho: [0,T] \times \R^m \to \R^{m\times d}$ are measurable
deterministic functions. Throughout we assume that there exists a $C\in \R_+$ such that for all $t\in[0,T]$ and
$x$, $x'\in \R^m$
\begin{itemize}
\item[(R1)] \qquad  $\begin{array}{ccc} |b(t,x) - b(t,x')| + |\rho(t,x) - \rho(t,x')| & \le & C |x-x'|, \\
  |b(t,x)| + |\rho(t,x)| & \le & C(1 +|x|). \end{array}$
\end{itemize}
We consider a derivative of the form $F(R_T)$, where $F: \R^m \to
\R$ is a bounded and measurable function. Note that at time $t$,
the expected payoff of $F(R_T)$, conditioned on $R_t = r$, is
given by $F(R^{t,r}_T)$, where $R^{t,r}$ is the solution of the
SDE
\begin{equation} \label{riskdyn}
R^{t,r}_s = r+ \int_t^s b(u,R^{t,r}_u) d u + \int_t^s \rho(u,R^{t,r}_u) d W_u, \quad s \in [t, T].
\end{equation}

Our correlated financial market consists of $k$ risky assets and
one non-risky asset. We use the non-risky asset as numeraire and
suppose that the prices of the risky assets in units of the
numeraire evolve according to the SDE
\[ dS^{i}_t = S^{i}_t (\alpha_i(t,R_t)  dt + \beta_i(t,R_t) dW_t), \quad i = 1,\ldots, k, \]
where $\alpha_i(t,r)$ is the $i$th component of a measurable and vector-valued map $\alpha: [0,T] \times \R^m
\to \R^k$ and $\beta_i(t,r)$ is the $i$th row of a measurable and matrix-valued map $\beta: [0,T] \times \R^m
\to \R^{k\times d}$. Notice that $W$ is the same $\IR^d$-dimensional Brownian motion as the one driving the
index process (\ref{riskpro}), and hence the correlation between the index and the tradable assets is determined by the matrices $\rho$ and $\beta$.

In order to exclude arbitrage opportunities in the financial market we assume $d \ge k$. For technical reasons we suppose that
\begin{enumerate}
\item[(M1)] $\alpha$ is bounded,
\item[(M2)] there exist constants $0 < \eps < K$ such that
$\eps I_k \le (\beta(t,r) \beta^*(t,r)) \le K I_k$ for all $(t,r)\in[0,T] \times \R^m$,
\end{enumerate}
where $\beta^*(t,r)$ is the transpose of $\beta(t,r)$, and $I_k$ is the $k$-dimensional unit matrix.

Before we proceed with the model description we will illustrate the range of possible applications by giving some examples of derivatives our model may apply to.
\begin{example}\label{ex.weather}
Weather derivatives are typical example of financial instruments
derived from non-tradable underlyings. One of the most common
types of weather derivatives are based on so-called {\em
accumulated heating degree days} (cHDD). The heating degree of a
day with average temperature $\tau$ in Celsius degrees is defined
as HDD $= \max\{0, 18 - \tau \}$, i.e. HDD describes the
(positive) difference between the average daily temperature
measured and the temperature above usually rooms are heated. The
cHDDs are defined as a moving average sum of HDDs over a fixed
time length, for instance a month. Real data shows the cHDD to be
almost lognormally distributed, and therefore they can be modelled
as geometric Brownian motions (see \cite{davis01}). This means
that in (\ref{riskpro}) we would have to choose $b(t, R_t) = \al_1
R_t$ and $\rho(t, R_t) = \al_2 R_t$, with $\al_1 \in \R$ and
$\al_2 \in \R\setminus\{0\}$ depending on the season. Tradable
assets that are more or less correlated with average temperatures
are for example electricity futures and natural gas futures.
\end{example}
The derivative explained in the next example is based on more than one underlying.
\begin{example}\label{ex.kerosene}
Spread options in general involve two or more underlying
structures (prices, indices, interest rates and many other
possible quantities), and measure the distance between them. We do
not go into details since spread options are well-known (see \cite{carmona} for an overview). For
simplicity we refer to a 2 dimensional example of \emph{Crack
spreads}.

Crack spreads consist in the simultaneous purchase or sale of
crude against the sale or purchase of refined petroleum products.
We concentrate on the kerosene crack spread, which pits crude oil
price (co) against kerosene price (ke). A company producing
kerosene (from crude oil) wishes to cover part of its risk arising
from a sudden boost of the crude oil price by buying kerosene
crack spreads. It thereby faces the problem that kerosene trading
is not done on a sufficiently liquid market to warrant a futures
contract or some other type of exchange-traded contract. So
derivative contracts of this type must be arranged on
over-the-counter basis.

Knowing that the price of heating oil (ho) is highly correlated
with the kerosene price - except during the Iraq war - crack
spreads themselves can be hedged by using heating oil futures.

We model prices in the following way, where the superscripts
represent the underlying products, \be
d R_t^{ke} &=& R^{ke}_t \left( b_1 dt + \gamma_2 dW^1_t+\gamma_3 dW^2_t+\gamma_4 dW^3_t \right)\\
d R_t^{co} &=& R^{co}_t \left( b_2 dt+\gamma_1 dW^1_t \right)\\
d S_t^{ho} &=& S^{ho}_t \left( b_3 dt+\beta_1 dW^1_t+\beta_2
dW^2_t \right), \ee where we assume that $b_1, b_2, b_3\in \IR$,
$\gamma_1, \gamma_2, \gamma_3, \gamma_4, \beta_1, \beta_2
\in\IR\backslash\{0\}$ and the correlation between heating oil and
kerosene is given by
$\sigma=(\gamma_2\beta_1+\gamma_3\beta_2)/\sqrt{(\gamma_2^2+\gamma_3^2+\gamma_4^2)(\beta_1^2+\beta_2^2)}$.


A European call on the spread is of the form
$\xi(R^{ke}_T,S_T^{co})=(R_T^{ke}-S_T^{co}-K)^+$, with $K$ being
the strike.
\end{example}

Throughout let $U$ be the exponential utility function with risk aversion coefficient $\eta > 0$, i.e.
\[ U(x) = - e^{-\eta x}. \]
In what follows let $(t,r)\in [0,T]\times \IR^m.$ By an {\em
investment strategy} we mean any predictable process $\lam =
(\lam^{i})_{1\le i\le k}$ with values in $\IR^k$ such that the
integral process $\int_0^t \lam^{i}_r \frac{dS^{i}_r}{S^{i}_r}$ is
defined for all $i \in\{1, \ldots, k\}$. We interpret $\lam^{i}$
as the value of the portfolio fraction invested in the $i$-th
asset.
Investing according to a strategy $\lam$ leads to a total gain due
to trading during the time interval $[t,s]$ which amounts to
$G^{\lam,t}_s = \sum_{i=1}^{k} \int_t^s \lam^{i}_u
\frac{dS^{i}}{S^{i}_u}$. We will denote by $G^{\lam,t,r}_s$ the
gain conditional on $R_t = r$.
\begin{remark}
As one can see the wealth process is given by
\[G^{\lambda,t,r}_s=\sum_{i=1}^k \int_t^s \lambda^i_u[\alpha_i(u,R_u^{t,r})du+\beta_i(u,R_u^{t,r})dW_u)],\]
and hence does {\em not} depend on the value of the correlated price process! This is a feature of our model that will later imply the indifference price at time $t$ to depend {\em only} on the value of the index process at a given time $t$.
\end{remark}
Let $\cA^{t,r}$ be the set of all strategies $\lam$ such that $E\int_t^T |\lam_s \beta(s, R^{t,r}_s)|^2 ds <
\infty$ and the family $\{e^{-\eta G^{\lam,t,r}_\tau}: \tau$ is a stopping time with values in $[t,T]\}$ is
uniformly integrable. If $\lam \in \cA^{t,r}$, then we say that $\lam$ is {\em admissible}. We use the same
admissibility criteria as in Section 2 in \cite{huimmu}, so that later we may invoke their results. The {\em
maximal expected utility} at time $T$, conditioned on the wealth to be $v$ at time $t$ and the index to satisfy
$R_t = r$, is defined by
\begin{equation} \label{cprb1}
V^0(t,v,r) = \sup\{ EU(v+G^{\lam,t,r}_T): \lam \in \cA^{t,r}\}.
\end{equation}
One can show that there exists a strategy $\pi$, called {\em optimal strategy}, such that
$EU(v+G^{\pi,t,r}_T) = V^0(v,t,r)$. The convexity of the utility functions implies that $\pi$ is a.s.\ unique on
$[t,T]$, and it follows from Theorem 7 in \cite{huimmu} that $\pi \in \cA^{t,r}$.

Suppose an investor is endowed with a derivative $F(R_T)$ and is
keeping it in his portfolio until maturity $T$. Then his maximal
expected utility is given by
\begin{equation} \label{cprb2}
V^F(t,v,r) = \sup\{ EU(v+G^{\lam,t,r}_T + F(R^{t,r}_T)): \lam \in \cA^{t,r}\}.
\end{equation}
Also in this case there exists an optimal strategy, denoted by $\widehat \pi$, that satisfies $EU(v+G^{\widehat
\pi,t,r}_T + F(R^{t,r})) = V^F(v,t,r)$.

The presence of the derivative $F(R_T)$ leads to a change in the
optimal strategy from $\pi$ to $\widehat \pi$. The difference
\[ \Delta = \widehat \pi - \pi \]
is needed in order to hedge, at least partially, the risk
associated with the derivative in the portfolio. We therefore call
$\Delta$ {\em derivative hedge}. In the following sections we
shall analyze by how much the optimal strategies change if a
derivative is added to the portfolio, and we aim at getting an
explicit expression for the derivative hedge $\Delta$.

One can easily show that for all $(t,r) \in [0,T]\times \R^m$ there exists a real number $p(t,r)$ such that for
all $v\in\IR$
\[ V^F(t,v -p(t,r),r) = V^0(t,v,r).\]
If an investor has to pay $p(t,r)$ for the derivative
$F(R^{t,r}_T)$, then he is indifferent between buying and not
buying the derivative. Therefore the number $p(t,r)$ is called
{\em indifference price} at time $t$ and level $r$.

It turns out that the derivative hedge $\Delta$ is closely related
to the indifference price of the derivative. The derivative either
diversifies or amplifies the risk exposure of the portfolio. The
difference between $\widehat \pi$ and $\pi$ measures the
diversifying impact of $F$. The price sensitivity, i.e. the
derivative of $p$ relative to the index evolution, is also a
measure of the diversification of $F$ (which will be called {\em
diversification pressure} of the derivative $F$). We will see that
the derivative hedge is indeed equal to the price sensitivity
multiplied with some correlation parameters.

The problem of finding the optimal strategies $\pi$ and $\widehat
\pi$ is a standard stochastic control problem. One can tackle it
by solving the related HJB equation, using a verification theorem
and proving a uniqueness result. This approach has been chosen for
example in \cite{05AIP}. Here, however, we prefer a stochastic
approach, using the fact that the stochastic control problem can
be solved by finding the solution of a backward stochastic
differential equation (BSDE). In the following section we briefly
recall the definition of a BSDE.
%
\section{Solving stochastic optimal control problems via BSDE}\label{controlviabsdes}
Let $\cH^2(\R^d)$ be the set of all $\R^d$-valued predictable
processes $\zeta$ such that $E\int_0^T |\zeta_t|^2 dt < \infty$,
and let $\cS^2(\R)$ be the set of all $\R$-valued predictable
processes $\delta$ satisfying $E\left(\sup_{s \in [0,T]}
|\delta_s|^2\right) < \infty$. By $\cS^\infty(\R)$ we denote the
set of all essentially bounded $\R$-valued predictable processes.
Let $\xi$ be $\cF_T$-measurable and $f$ a predictable mapping
defined on $\Omega\times[0,T]\times\R\times\R^d$ with values in
$\R$. A solution of the BSDE with {\em terminal condition} $\xi$
and {\em generator} $f$ is defined to be a pair of processes
$(Y,Z)\in \cS^2(\IR) \times \cH^2(\R^d)$ satisfying
\[ Y_t = \xi -\int_t^T Z_s dW_s + \int_t^T f(s,Y_s,Z_s) ds. \]
Let us now come back to our control problem of finding the optimal
investment strategy $\pi$ and $\widehat \pi$ respectively. It is
known that there exists a quadratic BSDE which solves these
control problems (see for example \cite{huimmu}). We first specify
the generator of the suitable BSDE, starting with $\hat{\pi}$.

Fix again $(t,r)\in[0,T]\times \R^m.$ Let $\theta(t,r) =
\beta^*(t,r)(\beta(t,r) \beta^*(t,r))^{-1} \alpha(t,r)$ and
$C(t,r)= \{x \beta(t,r): x \in \R^k\}$. Observe that our
assumptions imply that $\theta(t,r)$ is bounded. The distance of a
vector $z\in\R^d$ to the closed and convex set $C(t,r)$ will be
defined as $\dist(z,C(t,r)) = \min\{|z-u|: u \in C(t,r)\}$. Let
$f$ be the deterministic function
\[ f:[0,T] \times \R^m \times \R^d \to \R, \ (t,r,z) \mapsto z \theta(t,r)
+ \frac{1}{2\eta} |\theta(t,r)|^2 -\frac{\eta}{2}\dist^2(z + \frac{1}{\eta} \theta(t,r), C(t,r)).
\]
Since $d\geq k$, we have to find the orthogonal projection of the
$d-$dimensional vector $z$ to the linear space $C(t,r)$ of image
strategies. In \cite{huimmu} the set $C(t,r)$ is understood as
imposing restrictions on the investor when trading in the market
that happen to be convex in the setting given.

Notice that $f$ is differentiable in $z$ and satisfies the growth condition
\begin{equation*}
|f(t,r,z)|   \le c(1 + |z|^2)  \quad \textrm{ a.s.} 
\end{equation*}
with some $c \in \R_+$. The growth condition guarantees that there
exists a unique solution $(\widehat Y^{t,r}, \widehat Z^{t,r})\in
\cS^\infty(\R) \otimes \cH^2(\R^d)$ of the BSDE
\begin{equation}\label{bsdemit}
\widehat Y^{t,r}_s = F(R^{t,r}_T) -\int_s^T \widehat Z^{t,r}_u dW_u - \int_s^T f(u,R^{t,r}_u,\widehat Z^{t,r}_u) du,
\quad s\in[t,T],
\end{equation}
(see Theorem 2.3 and 2.6. in \cite{koby}). Notice that the
terminal condition of the BSDE stems from a standard forward SDE.
The system of equations consisting of (\ref{riskdyn}) and
(\ref{bsdemit}) is often called {\em forward-backward stochastic
differential equation} (FBSDE).

The conditional maximal expected wealth, or in other words the value function of our stochastic control problem,
is equal to the utility of the starting point of the BSDE, i.e.
\[ V^F(t,v,r) = -e^{-\eta(v - \widehat Y^{t,r}_t)}\]
(see Theorem 7 in \cite{huimmu}). Moreover we can reconstruct the optimal strategy $\widehat \pi$ starting from
$\widehat Z$. To this end denote by $\Pi_{C(t,r)} (z)$ the projection of a vector $z \in \R^d$ onto the linear
subspace $C(t,r)$. If $R_t = r$, then the optimal strategy $\widehat \pi_t$ on $[t,T]$ satisfies
\begin{equation}\label{markovstra}
 \widehat \pi_s \beta(s,R^{t,r}_s) = \Pi_{C(s,R^{t,r}_s)} [\widehat Z^{t,r}_s +
\frac1\eta \theta(s,R^{t,r}_s)], \quad s\in[t,T].
\end{equation}
The last statement follows equally from Theorem 7 in \cite{huimmu}. 

Analogously, let $(Y^{t,r}, Z^{t,r})$ be the solution of 
\begin{equation}\label{bsdeohne}
Y^{t,r}_s = -\int_s^T Z^{t,r}_u dW_u - \int_s^T
f(u,R^{t,r}_u,Z^{t,r}_u) du, \quad s\in[t,T],
\end{equation}
which represents a stochastic control problem as above, just
without the derivative as terminal condition i.e. the derivative is not in the portfolio. In this case the maximal expected utility verifies 
$$V^0(t,v,r) = -e^{-\eta(v - Y^{t,r}_t)},$$ and the optimal strategy
$\pi$ on $[t,T]$ satisfies
\begin{equation}\label{markovstra2}
\pi_s \beta(s,R^{t,r}_s) = \Pi_{C(s,R^{t,r}_s)} [Z^{t,r}_s + \frac1\eta \theta(s,R^{t,r}_s)], \quad s\in[t,T].
\end{equation}
Since $\Pi_{C(s,R^{t,r}_s)}$ is a linear operator, the derivative hedge is
given by the explicit formula
\[ \Delta_s \beta(s,R^{t,r}_s) = \Pi_{C(s,R^{t,r}_s)} [\widehat Z^{t,r}_s - Z^{t,r}_s], \]
which will be further determined in the subsequent sections.
\section{The Markov property of the indifference prices}\label{markovofprice}
In this section we will establish the Markov property of the
indifference prices. This will follow from the fact that the
solutions of the BSDEs (\ref{bsdemit}) and (\ref{bsdeohne}) are
deterministic functions of time and the underlying. To give the
precise statement we need to introduce the following
$\sigma$-algebras. Fixing $t\in[0,T]$, we denote by $\cD^m$ the
$\sigma$-algebra generated by the functions $r\mapsto E[\int_t^T
\phi(s,R^{t,r}_s)ds]$, where $t\in[0,T]$ and $\phi$ is a bounded continuous
$\IR-$valued function.

Moreover we assume that the mapping $(t,r) \mapsto \theta(t,r)$ be
Lipschitz continuous in $r$, noting that due to (M1) and (M2) this
is guaranteed if $\beta$ and $\alpha$ are Lipschitz continuous.
\begin{lemma} \label{markovappl}
There exist $\cB[0,T] \otimes \cD^m$-measurable deterministic functions $u$ and $\widehat u: [0,T] \times \R^m
\to \R$ such that
\begin{eqnarray*}
Y^{t,r}_s &=& u(s, R^{t,r}_s) \quad \textrm{ and }  \quad \widehat Y^{t,r}_s = \widehat u(s, R^{t,r}_s),
\end{eqnarray*}
for $P\otimes \lambda$-a.a.\ $(\om, s) \in \Omega \times [t,T]$.
\end{lemma}
\begin{proof}
The generator function $f$ is a polynomial of the components of
$z$ of at most second degree. This implies, together with the
assumption that $\theta$ is Lipschitz continuous in $r$, that
there exist functions $f_n: [0,T]\times \R^m \times \R^d \to \R$,
globally Lipschitz continuous in $z$, such that for all compact
sets $K \subset \R^m \times \R^d$ the sequence $f_n$ converges to
$f$ uniformly on $[0,T] \times K$. Thus the statement follows from
Theorem \ref{markov}.
\end{proof}
Lemma \ref{markovappl} immediately implies that there exists a
nice version of the indifference price $p$ as a function of
$(t,r)$.
\begin{theo}\label{onlyrisk}
There exists a $\cB[0,T] \otimes \cD^m$-measurable deterministic
function $p: [0,T] \times \R^m \to \R$ such that for all $v\in\IR,
(t,r)\in[0,T]\times\R^m$
\begin{equation}\label{dynindi}
V^F(t,v -p(t,r), r) = V^0(t,v,r).
\end{equation}
\end{theo}
\begin{proof}
Let $v\in\IR, (t,r)\in[0,T]\times\R^m$ be given. Recall that
$V^F(v,t,r) = -e^{-\eta(v - \widehat Y^{t,r}_t)}$ and $V^0(v,t,r)
= -e^{-\eta(v - Y^{t,r}_t)}$. Then put $p(t,r) = u(t,r) - \widehat
u(t,r)$, where $u$ and $\widehat u$ are given from Lemma
\ref{markovappl}.
\end{proof}
In the remainder the function $p$ is always assumed to be
measurable in both $t$ and $r$. In fact it inherits this property
from the functions $u$ and $\hat{u}.$

We now turn to an explicit description of the optimal strategies,
and in particular their difference, the derivative hedge. These
will be derived from the BSDE solutions of the preceding section.
We start by noting that similarly to the indifference price the
optimal strategies only depend on the time and the index process
$R$.
\begin{theo}\label{onlyrisk2}
There exist $\cB[0,T] \otimes \cD^m$-measurable deterministic
functions $\nu$ and $\widehat \nu$, defined on $ [0,T] \times
\R^m$ and taking values in $\R^d$ such that for
$(t,r)\in[0,T]\times \R^m$, the optimal strategies, conditioned on
$R_t = r$, are given by $\pi_s = \nu(s,R^{0,r}_t)$ and $\widehat
\pi_s = \widehat \nu(s,R^{t,r}_s)$ for all $s\in[t,T]$.
\end{theo}
\begin{proof}
Fix $(t,r)\in[0,T]\times \R^m.$ Theorem \ref{markov} implies that
there exist $\cB[0,T] \otimes \cD^m$-measurable deterministic
functions $v$ and $\widehat v$ mapping  $[0,T] \times \R^m$ to
$\R^m$ such that for all $s\in[t,T]$
\[ Z^{t,r}_s = v(s,R^{t,r}_s) \rho(s,R^{t,r}_s) \quad \textrm{ and }
\quad \widehat Z^{t,r}_s = \widehat v(s,R^{t,r}_s) \rho(s,R^{t,r}_s). \] Now let $\gamma(t,r) = \Pi_{C(t,r)}
[v(t,r) \rho(t,r) + \frac{1}{\eta} \theta(t,r)]$ and $\widehat \gamma(t,r) = \Pi_{C(t,r)} [\widehat v(t,r)
\rho(t,r) + \frac{1}{\eta} \theta(t,r)]$. Then, by (\ref{markovstra}) and (\ref{markovstra2}), the optimal
strategies conditioned on
$R_t = r$ satisfy 
\[ \widehat \pi_s \beta(s,R^{t,r}_s) = \widehat \gamma(s,R^{t,r}_s) \quad \textrm{ and }
\quad \pi_s \beta(s,R^{t,r}_s) = \gamma(s,R^{t,r}_s), \] for all
$s \in [t,T]$. Since the rank of $\beta(t,r)$ is $k$, then both
$\widehat \nu(t,r) = \widehat \gamma(t,r) \beta^*(t,r) (\beta(t,r)
\beta^*(t,r))^{-1}$ and $\nu(t,r) = \gamma(t,r) \beta^*(t,r)
(\beta(t,r) \beta^*(t,r))^{-1}$ are well defined. Then uniqueness
of $\pi$ and $\widehat \pi$ yields the result.
\end{proof}
\begin{remark}
Theorem \ref{onlyrisk2} implies that the optimal strategies are
the so-called \emph{Markov controls}.
\end{remark}
We close this section by noting that Theorem \ref{onlyrisk}
implies a dynamic principle for the indifference price. Abbreviate
$\cA = \cA^{0, r}$ for some $r\in\R^m.$ For any stopping time
$\tau \le T$ and $\cF_\tau$-measurable random variable $G_\tau$
let $V^F(\tau, G_\tau) = \textrm{esssup}_{} \{E[U(G_\tau +
G^{\lam,\tau}_T + F(R^{0,r}_T)|\cF_\tau]: \lam \in \cA \}$. Similarly we define
$V^0(\tau, G_\tau)$.
\begin{corollary} We have
\[ V^F(\tau, G_\tau - p(\tau,R^{0,r}_\tau)) = V^0(\tau, G_\tau). \]
\end{corollary}
\begin{proof}
As is shown in Prop. 9 in \cite{huimmu}, the value function  $V^F$
satisfies the dynamic principle
\[ V^F(\tau, G_\tau- p(\tau,R^{0,r}_\tau)) = U(G_\tau - p(\tau, R^{0,r}_\tau) - \widehat Y^{0,r}_\tau). \]
Since $p(\tau,R^{0,r}_\tau)= Y^{\tau, R^{0,r}_\tau}_\tau - \widehat Y^{\tau, R^{0,r}_\tau}_\tau = Y^{0,r}_\tau - \widehat Y^{0,r}_\tau$ we obtain
$V^F(\tau, G_\tau- p(\tau,R^{0,r}_\tau)) = U(G_\tau -  Y^{0,r}_\tau) = V^0(\tau, G_\tau)$.
\end{proof}
\section{Differentiable indifference prices and explicit hedging strategies}\label{explicithedge}
If we impose stronger conditions on the coefficients of the index
process $R$ and the function $F$, then we can show that the price
function $p$ is differentiable in $r$, and we can obtain an
explicit representation of the derivative hedge in terms of the
price gradient. To this end we need to introduce the following
class of functions.
\begin{defi}\label{bfb}
Let $n$, $p \ge 1$. We denote by $\bf B^{n \times p}$ the set of
all functions $h:[0,T] \times \R^m \to \R^{n \times p}$, $(t,x)
\mapsto h(t,x)$, differentiable in $x$, for which there exists a
constant $C >0$ such that \linebreak $\sup_{(t,x)\in[0,T]
\times\IR^m}  \sum_{i=1}^m \left| \frac{\partial h(t,x)}{\partial
x_i} \right| \le C$, for all $t\in[0,T]$ we have $\sup_{x \in\R^m}
\frac{|h(t,x)|}{1+|x|} \le C$, and $x \mapsto \frac{\partial
h(t,x)}{\partial x}$ is Lipschitz continuous with Lipschitz
constant $C$.
\end{defi}
We will assume that the coefficients of the index diffusion satisfy in addition to (R1)
\begin{itemize}
\item[(R2)]\label{riskcoeff}
$\rho \in \bf B^{m \times d}$, $b \in \bf B^{m\times 1}$, and
\item[(R3)] $F$ is a bounded and twice differentiable function such that\[ \nabla F \cdot \rho \in \textrm{\bf{B}}^{1 \times d}
\textrm{ and }\sum_{i=1}^m b_i(t,r) \frac{\partial}{\partial r_i} F(r) +
\frac12 \sum_{i, j=1}^m [\rho\rho^{*}]_{ij}(t,r) \frac{\partial^2}{\partial r_i \partial r_j} F(r) \in \textrm{\bf B}^{1\times 1}.\]
\end{itemize}
The next result guarantees Lipschitz continuity and
differentiability of the functions $u$ and $\widehat u$ obtained
from Theorem \ref{markovappl}.

\begin{theo}\label{diffbsdeappl}
Suppose that (R1), (R2) and (R3) are satisfied. Besides, suppose that the volatility matrix $\beta$ and the drift density $\alpha$ are bounded, Lipschitz continuous in $r$, differentiable in $r$ and that for all $1\le i\le k$, $1 \le j \le d$ the derivatives $\nabla_r \beta_{ij}$ and $\nabla_r \alpha_i$ are also Lipschitz continuous in $r$. Then the functions $u$ and $\widehat u$ are
Lipschitz continuous in $r$, and continuously differentiable in $r$.
\end{theo}
\begin{proof}
The theorem follows from Lemma \ref{300307-2} and Theorem \ref{markov2} in Section \ref{sec6}. All we have to show at this stage is that the assumptions of both results are satisfied. We only show Conditions (\ref{lipinx}) and (\ref{lipinnablax}), since the remaining ones are easily seen to be fulfilled. 

Notice that the conditions on $\alpha$ and $\beta$ imply that $\theta$ is
differentiable in $r$, and that $\theta$ and $\nabla_r \theta$ are globally Lipschitz continuous in $r$. Moreover, since $\theta$ is bounded, $\theta^2$ is Lipschitz continuous in $r$, too. Recalling the definition of the generator $f$, note further that 
\be
& & |\dist^2(z + \frac{1}{\eta} \theta(t,r),C(t,r))  - \dist^2(z + \frac{1}{\eta} \theta(t,r'),C(t,r)) | \\
&\le& 
2(\frac{1}{\eta}\|\theta\|_\infty + |z|) \  |\dist(z + \frac{1}{\eta} \theta(t,r),C(t,r))  - \dist(z + \frac{1}{\eta} \theta(t,r'),C(t,r)) |\\
&\le& 
2(\frac{1}{\eta}\|\theta\|_\infty + |z|) \  \left| |z + \frac{1}{\eta} \theta(t,r) - \Pi_{C(t,r)}(z + \frac{1}{\eta} \theta(t,r))|  - |z + \frac{1}{\eta} \theta(t,r') - \Pi_{C(t,r)}(z + \frac{1}{\eta} \theta(t,r'))| \right|\\
&\le& 
4(\frac{1}{\eta}\|\theta\|_\infty + |z|) \  \frac{1}{\eta} \left| \theta(t,r) - \theta(t,r') \right|,  \qquad t\in[0,T], r, r' \in \R^m, z \in \R^d. 
\ee
This shows that there exists a constant $K\in \R_+$ such that $|f(t,r,z) - f(t,r',z)|\le K(1 + |z|) |r -r'|$, and hence Assumption (\ref{lipinx}) of Lemma \ref{300307-2} is satisfied. 

Observe that $\Pi_{C(t,r)}(y) = y \beta^* (\beta \beta^*)^{-1} \beta(t,r)$ for all $y\in\R^k$, and hence the mapping $r \mapsto \Pi_{C(t,r)}(z + \frac{1}{\eta} \theta(t,r))$ is differentiable. Consequently, also $f$ is differentiable and for $t\in[0,T]$, $r \in \R^m$ and $z \in \R^d$ we have
\be
\nabla_r f(t,r,z) &=& z \nabla_r \theta(t,r) + \frac{1}{\eta} \theta(t,r) \nabla_r \theta(t,r) \\
& &- \eta \left( z + \frac{1}{\eta} \theta(t,r) - \Pi_{C(t,r)}(z + \frac{1}{\eta} \theta(t,r)) \right) \left( \frac{1}{\eta} \nabla_r \theta(t,r) - \nabla_r \Pi_{C(t,r)}(z + \frac{1}{\eta} \theta(t,r)) \right)
\ee
By using that $\theta$, $\nabla_r \theta$, $\beta$ and $\nabla_r \beta$ are Lipschitz continuous and bounded, it is straightforward to show that for all $t\in[0,T]$, $r,r' \in \R^m$ and $z,z' \in \R^d$ we have 
\be
|\nabla_r f(t,r,z) - \nabla_r f(t,r',z')|\le K(1 + |z| + |z'|) (|r -r'|+|z-z'|),
\ee
and hence the generator satisfies Assumption (\ref{lipinnablax}) of Theorem \ref{markov2}.

Now Lemma \ref{300307-2} yields the Lipschitz continuity in $r$ of the functions $u$ and $\widehat u$. Theorem
\ref{markov2} implies the differentiability of $\widehat Y^{t,r}$ and $Y^{t,r}$ with respect to $r$, and hence
also of $u$ and $\widehat u$. 
\end{proof}
As an immediate consequence we obtain smoothness of the indifference price function.
\begin{corollary}\label{diffbarkeit}
Suppose that the assumptions of Theorem \ref{diffbsdeappl} are
satisfied. Then the indifference price function $p$ is
continuously differentiable in $r$.
\end{corollary}
Having shown smoothness of the indifference price, we can finally
derive an explicit formula for the derivative hedge in terms of
the price gradient. To this end we denote the \emph{conditional
derivative hedge} by $\Delta(t,r) = \widehat \nu(t,r) - \nu(t,r),
(t,r)\in[0,T]\times \R^m$.
\begin{theo}\label{noconstraints}
Under the assumptions of Theorem \ref{diffbsdeappl}, and with the
notation of Section \ref{controlviabsdes}, the derivative hedge
satisfies
\begin{equation}\label{optstr}
\Delta(t,r)= - \nabla_r p(t,r)
\rho(t,r) \beta^*(t,r) (\beta(t,r) \beta^*(t,r))^{-1},\quad
(t,r)\in[0,T]\times \R^m.
\end{equation}
\end{theo}
\begin{remark} Note that Theorem \ref{noconstraints} implies that the derivative hedge at time $t$ depends only on $R_t$.
\end{remark}
\begin{proof}[ Proof of Theorem \ref{noconstraints}]
Note that $C(t,r)$ is a linear subspace of $\R^d$ for all
$(t,r)\in[0,T] \times \R^m$. Therefore, the projection operator
$\Pi_{C(t,r)}$ is linear and hence \be
\Delta(t,r) &=& \left(\widehat \gamma(t,r) - \gamma(t,r)\right) \beta^*(t,r) (\beta(t,r) \beta^*(t,r))^{-1} \\
&=& \left(\Pi_{C(t,r)}[\widehat Z^{t,r}_t+ \frac{1}{\eta} \theta(t,r)] - \Pi_{C(t,r)} [Z^{t,r}_t+ \frac{1}{\eta} \theta(t,r)]\right) \beta^*(t,r) (\beta(t,r) \beta^*(t,r))^{-1} \\
&=& \left(\Pi_{C(t,r)}[\widehat Z^{t,r}_t -Z^{t,r}_t]\right) \beta^*(t,r) (\beta(t,r) \beta^*(t,r))^{-1}.\\
\ee
It follows from Theorem \ref{markov2} that $\widehat Z^{t,r}_t -Z^{t,r}_t = \left(\nabla_r \widehat u(t,r) - \nabla_r u(t,r)\right) \rho(t,r) = - \nabla_r p(t,r) \rho(t,r)$, and hence we obtain the result.
\end{proof}
If the market consists of only one risky asset, then the optimal strategy simplifies to the following formula.
\begin{corollary}
Let $k=1$. Then the derivative hedge is given by
\[ \Delta(t,r) = - \frac{\langle \beta(t,r), \nabla_r p(t,r) \rho(t,r)\rangle }{|\beta(t,r)|^2}= - \frac{\sum_{i= 1}^d
\beta_i(t,r)   \sum_{j=1}^m \frac{\partial}{\partial r_j}p(t,r)
\rho_{ji}(t,r) }{ \sum_{i= 1}^d \beta_i^2(t,r) },\quad
(t,r)\in[0,T]\times\R^m.
\]
\end{corollary}
\begin{proof}
Fix $(t,r)\in [0,T]\times \R^m.$ Note that $C(t,r) = \{x
\beta(t,r): x \in \R\}$ is a one-dimensional subspace of $\R^d$.
For all $z = (z_i)_{1\le i \le d} \in \R^d$ let $g(z) =
\frac{\langle \beta(r,t), z\rangle}{|\beta(t,r)|^2} =
\frac{\sum_{i= 1}^d \beta_i(t,r) \ z_i }{\sum_{i= 1}^d
\beta_i^2(t,r)}$. Then $g(z)\beta(t,r)$ is the orthogonal
projection of $z$ onto $C(t,r)$. Thus Theorem \ref{noconstraints}
yields that $\Delta(t,r) = - g(\nabla_r p(t,r)\rho(t,r))$.
\end{proof}
\begin{remark}$\phantom{12}$\\
1) Suppose the derivative $F(R_T)$ is traded on an exchange. By
pretending the price observed is approximately equal to an
indifference price, the hedging formula (\ref{optstr}) provides a
very simple tool for hedging the derivative. Notice that the risk
aversion coefficient $\eta$ does not appear explicitly in
(\ref{optstr}). \\
2) If $k=d$ and the matrices $\beta(t,r)$ are all invertible, then our financial market is complete and the
derivative $F(R_T)$ can be fully replicated. Moreover the derivative hedge satisfies
\[ \Delta(t,r) = - \nabla_r p(t,r) \rho(t,r)\beta^{-1}(t,r). \]
If $S$ is chosen to be the index, i.e. $R = S$, then we obtain
$\Delta = \left(\begin{array}{ccc} \frac{\partial p}{\partial
s_1}(t,r) S^1, \cdots, \frac{\partial p}{\partial s_k}(t,r)
S^k\end{array}\right)$. Moreover, the number of shares to invest
into asset $i$ is given by $\frac{\Delta^{i}(t,r)}{S^{i}(t,r)} =
\frac{\partial p}{\partial s_i}$. Thus $\Delta$ coincides with the
classical 'delta hedge'.
\end{remark}
\begin{example}
As in Example \ref{ex.weather} suppose that $R$ is the moving
average cHDD process modelled as a geometric Brownian motion, and
assume that there exists one tradable correlated risky asset. More
precisely let $d=2$, $k=m=1$, $\rho = \left(
\begin{array}{cc} \al_2 & 0 \end{array}\right)$, $\beta = \left(\begin{array}{cc}\beta_1 & \beta_2 \end{array}
\right)$ with $\al_2, \beta_1, \beta_2 \in \IR\setminus \{0\}$. Then
\[ \Delta(t,r) = - \al_2 \frac{\partial p(t,r)}{\partial r} \frac{\beta_1}{\beta_1^2 + \beta_2^2}. \]
\end{example}
\begin{example} \label{ex.ker2}Applying our results to Example
\ref{ex.kerosene}, we have to take $m=2$, $k=2$ and $d=3$. Hence
\begin{displaymath}
\rho=\left(
\begin{array}{ccc}
\gamma_1 & 0 &0\\
\gamma_2 &\gamma_3 &\gamma_4
\end{array}
\right),
\qquad
\beta=\left(
\begin{array}{ccc}
\gamma_1 & 0 &0\\
\beta_1 &\beta_2 &0
\end{array}
\right),
\qquad
\beta^*(\beta\beta^*)^{-1}=\frac{1}{\gamma_1 \beta_2}\left(
\begin{array}{cc}
\beta_2 & 0 \\
-\beta_1 & \gamma_1 \\
0 & 0
\end{array}
\right).
\end{displaymath}
With a simple minimum square calculation we compute
\[
\Pi_{C(t,r)}[\nabla_r p(t,r) \rho(t,r)]=
\left(
\begin{array}{ccc}
\gamma_1\frac{\partial}{\partial
{r_1}}p(t,r)+\gamma_2\frac{\partial}{\partial{r_2}}p(t,r) & \
\gamma_3 \frac{\partial}{\partial{r_2}} p(t,r) & \ 0
\end{array}\right)
\]
Equation (\ref{optstr}) applied to our example produces the
following Delta hedge for $(t,r)\in[0,T]\times \R^2$
\begin{displaymath}
\Delta(t,r)=
\left(
\begin{array}{cc}
-\frac{\partial}{\partial{r_1}} p(t,r)+(\frac{\beta_1
\gamma_3}{\gamma_1
\beta_2}-\frac{\gamma_2}{\gamma_1})\frac{\partial}{\partial{r_2}}
p(t,r) & \
 -\frac{\gamma_3}{\beta_2} \frac{\partial}{\partial{r_2}} p(t,r)
\end{array}\right)
\end{displaymath}
where $r_1$ represents the crude oil and $r_2$ the kerosene
variable. If $\gamma_4=0$ then we have a perfect hedge and if
$\gamma_3=0$, then the price of heating oil doesn't play a role in
the hedge, as one would expect.
\end{example}
\section{Pricing by marginal utility}\label{section:MUP}
Suppose there is no exchange and the derivative $F(R_T)$ is sold
over-the-counter. What is a reasonable price a seller could ask
for the derivative? The indifference price seems to be a natural
candidate, though it has the disadvantage that the price of a
single derivative depends on the total quantity sold, i.e. the
indifference price is non-linear. For instance the indifference
price of $2\times F(R_T)$ does not equal twice the indifference
price of $F(R_T)$. In order to obtain a linear version one may
take the limit of the indifference price as the quantity converges
to $0$. The object thus derived is the indifference price for a
vanishing amount of derivatives, and it is therefore called {\em
marginal utility price} (MUP). Having to pay the MUP for each
derivative an investor is indifferent between buying and not
buying an infinitesimal amount of the derivative.

We continue requiring (R1)-(R3) to be satisfied. We update the
notation and, for $q\in\R$ and $(t,r)\in[0,T]\times \R^m$ define
by $p(t,r,q)$ the indifference price of $q$ units of
$F(R^{t,r}_T)$, i.e. $p(t,r,q)$ is the unique real satisfying
\[ \sup_\lam\{ EU(v+G^{\lam,t,r}_T + q F(R^{t,r}_T) - p(t,r,q)\} = \sup_\lam\{ EU(v+G^{\lam,t,r}_T)\}. \]
The price of one unit is equal to $\frac{p(t,r,q)}{q}, (q\not= 0)$, and the MUP is defined by
\[\textrm{MUP}(t,r) = \frac{\partial}{\partial q} p(t,r,q) \big|_{q=0}.  \]
Recall that $p(t,r,q) = Y^{t,r}_t - \widehat Y^{t,r,q}_t$, where
$(\widehat Y^{t,r,q},\widehat Z^{t,r,q})$ is the solution of the
BSDE
\begin{equation*}
\widehat Y^{t,r,q}_s = q F(R^{t,r}_T) -\int_s^T \widehat Z^{t,r,q}_u dW_u - \int_s^T f(u,R^{t,r}_u,\widehat
Z^{t,r,q}_u) du, \quad s\in[t,T].
\end{equation*}

Naming $\xi(q)=qF(R^{t,r}_T)$, then clearly $\xi(q)$ is a globally
bounded differentiable Lipschitz function (with bounded
derivatives). The boundedness of $\xi$ is trivial since $F$ is
bounded and we are only interested in the differentiability of the
process with relation to $q$ in a neighborhood of zero. And so,
due to the boundedness of $F$ and the quadratic growth hypothesis
for $f$ the conditions of Theorem \ref{diffbsde.xi} are satisfied.
Hence, the  process $\widehat{Y}^{t,r,q}$ is continuous in $t$ and
continuously differentiable in $q$.

Writing the BSDE differentiated with respect to $q$ gives
\begin{equation*}
\frac{\partial }{\partial q} \widehat Y^{t,r,q}_s = F(R^{t,r}_T) -
\int_s^T \frac{\partial}{\partial q} \widehat Z^{t,r,q}_u dW_u -
\int_s^T \nabla_z f(u,R^{t,r}_u,\widehat Z^{t,r,q}_u)
\frac{\partial}{\partial q} \widehat Z^{t,r,q}_u du, \quad
s\in[t,T].
\end{equation*}
Setting $q = 0$ and renaming the processes for ease of notation we
obtain \ben U^{t,r}_s = F(R^{t,r}_T) -\int_s^T V_s d W_s -
\int_s^T \nabla_z f(s,R^{t,r}_s,Z^{t,r}_s)\cdot V_s ds.\een As an
end product of these calculations we obtain the following explicit
formula for the (MUP) of our derivative.
\begin{theo}
The explicit formula for the Marginal Utility Price of the
derivative $F(R_T)$ is given by
\[ MUP(t,r) =
U^{t,r}_t, \] where $U^{t,r}_t$ is  the first component of
the solution pair of the BSDE \ben\label{MUP} U^{t,r}_s =
F(R^{t,r}_T) -\int_s^T V_s d W_s - \int_s^T \nabla_z
f(s,R^{t,r}_s,Z^{t,r}_s)\cdot V_s ds.\een
\end{theo}
\begin{remark}
Note that by performing a Girsanov change of probability measure
to the one making the process $\tilde{W} = W + \int_0^\cdot
\nabla_z f(s, R_s^{t,r}, Z_s^{t,r}) ds$ a Brownian motion, solving
(\ref{MUP}) reduces to taking conditional expectations with
respect to the underlying filtration. Hence, denoting by
$\cE(\cdot)$ the stochastic exponential operator, we can represent
the marginal utility price explicitly by the following expression
\be MUP(t,r) &=& E\Big[\cE\Big(\int_0^\cdot \nabla_z
f(s,R^{t,r}_s,Z^{t,r}_s) dW_s\Big)_t^T F(R^{t,r}_T) \Big].
\ee
\end{remark}

\section{Some mathematical tools: smoothness of quadratic FBSDE} \label{sec6}
%
%
%
%
\subsection{Moment estimates for BSDE with random Lipschitz condition}
In the following we provide moment estimates for BSDE with
generators that satisfy Lipschitz conditions with random bounds
for the slopes. More precisely, we assume that for our generator
$f:\Omega \times [0,T] \times \IR^d \to \R$ there exists an
$\R_+$-valued predictable process $H$ such that for all $(\om,t,z)
\in \Omega \times [0,T] \times \IR^d$ we have
\begin{equation} \label{bmolipschitz}
|f(\om,t,z) - f(\om,t,z')| \le H_t |z-z'|.
\end{equation}
We will assume that $H$ is such that the stochastic integral
$\int_0^\cdot H dB$ with respect to a Brownian motion $B$ is a
so-called $\bmo$ martingale. Recall that $\int_0^\cdot H dB$
is a $\bmo$ martingale (we also say it belongs to \bmo) if and
only if there exists a constant $C \in \IR_+$ independent of
$\omega$ such that for all stopping times $\tau$ with values in
$[0,T]$ we have
\begin{equation}\label{bmo}
E\left[\int_\tau^T H_s^2 ds \Big|\cF_\tau \right] \le C, \quad \textrm{ a.s.}
\end{equation}
We refer to \cite{kazamaki} for basic information about BMO
martingales. We will abuse the definition and refer to the
smallest $C\in \R_+$ that satisfies inequality (\ref{bmo}) as the
\bmo \ norm of $H$.

Throughout let $W$ be a $d$-dimensional Brownian motion. Consider
the BSDE
\begin{equation} \label{bsde-lin}
Y_t = \xi - \int_t^T Z_s dW_s + \int_t^T f(s,Z_s) d s, \quad 0\le
t\le T,
\end{equation}
where $\xi$ is a bounded $\cF_T$-measurable random variable, and
$f$ satisfies (\ref{bmolipschitz}) relative to a predictable $H$
with finite $\bmo$ norm.

We refer to \cite{bricon} for sufficient criteria for the existence of solutions of such BSDEs.

The moment estimate we shall give next will be needed later for establishing smoothness of the solution of the
quadratic BSDE with respect to the parameters the terminal condition depends on.
\begin{lemma} \label{apriori-quadra}
Suppose that for all $\beta \ge1$ we have $\int_0^T |f(s,0)| d s
\in L^\beta(P)$. Let $p >1$. Then there exist constants $q>1$ and
$C > 0$, depending only on $p$, $T$,  and the
BMO-norm of $H$, such that we have \be E \Big[\sup_{t\in[0,T]}
|Y_t|^{2p} \Big]+ E \left[ \left(\int_0^T |Z_s|^2 d s \right)^{p}
\right]
&\leq& C \, \left(  E\Big[\,|\xi|^{2p q} +(\int_0^T |f(s,0)| d s)^{2pq} \Big]\right)^{\frac{1}{q}}. \ee
\end{lemma}
\begin{proof}
This follows from Corollary 3.4 in \cite{bricon} by a straightforward generalization to the multidimensional case
considered here. It can also be shown with the method used in the proof of Theorem 5.1 in \cite{animre}.
\end{proof}
\subsection{Differentiability of quadratic FBSDE}\label{subsec:diff.fbsde}
Consider now a FBSDE of the form \ben \label{fbsde}
\begin{array}{ccl}X^{x}_s & = & x+\int_0^t b(s,X^{x}_s) d s+\int_0^t \rho(s,X^{x}_s) d W_s, \\
Y^x_s & = & F(X^x_T) - \int_t^T Z^x_s dW_s + \int_t^T f(s,X^{x}_s, Z^x_s) ds,
\end{array}
\een where $b:[0,T]\times\R^m\to\R^m$ and $\rho:[0,T]\times \R^m
\to \R^{m\times d}$ and $W$ is the $d-$dimensional Brownian motion
of the preceding subsection. Note that $\rho$ is a $n\times d$
matrix. We will denote its transpose by $\rho^*$. The generator of
the backward part is assumed to be a $\cP(\cF_t) \otimes \cB(\R^m)
\otimes \cB(\R^d)$-measurable process $f: \Omega \times [0,T]
\times \R^m \times \R^{d} \to \R$ such that there exists a
constant $M\in\R_+$ such that for all $(t,x,z) \in
[0,T]\times\R^m\times \R^d$ we have
\begin{equation}\label{subquadra}
|f(t,x,z)| \le M(1 + |z|^2)  \quad \textrm{a.s.}
\end{equation}
Here $\cP(\cF_t)$ denotes the $\sigma$-field of predictable sets
with respect to the filtration $(\cF_t).$ Moreover we assume that
\begin{eqnarray}\label{lipabl}
\begin{array}{c}f \textrm{ is differentiable in $x$ and $z$ and} \\
|\nabla_z f(t,x,z)| \le M(1+|z|) \quad \textrm{ for all }
(t,x,z)\in[0,T]\times\R^m\times \R^d \textrm{ a.s.}
\end{array}
\end{eqnarray}
We will give sufficient conditions for the process $Y^{x}$ in the
solution of the FBSDE (\ref{fbsde}) to be differentiable in $x$. A
further assumption we need is that the coefficients of the forward
equation belong to the function space $\bf B^{m \times d}$ and
$\bf B^{m \times 1}$ respectively (see Definition \ref{bfb}). To
simplify notation, to the pair $(b, \rho)$ of coefficient
functions we associate the second order differential operator $\cL
= \sum_{i=1}^m b_i(\cdot) \frac{\partial}{\partial x_i} + \frac12
\sum_{i, j=1}^m [\rho\rho^*]_{ij}(\cdot)
\frac{\partial^2}{\partial x_i
\partial x_j}$.
%
%

We will assume that the coefficients of the forward equation (\ref{fbsde}) satisfy
\begin{itemize}
\item[({D}1)]\label{fbforward}
$\rho\in \bf B^{m \times d}$, $b \in \bf B^{m\times 1}$,
\end{itemize}
and that
\begin{itemize}
\item[({D}2)]\label{fbterminal} $F:\R^m\to \R$ is a twice
differentiable function such that $\nabla F \cdot \rho \in \bf
B^{1 \times d}$ and $\cL F \in \bf B^{1\times 1}$.
\end{itemize}
It is known that the conditions (D1) and (D2) ensure that $X^{x}$
is differentiable in $x$ and the difference quotients can be
nicely controlled. For the convenience of the reader we quote a
standard result which will be needed later. Denote by $e_i$ the
unit vector in $\IR^m$ in the direction of coordinate $i$, $1\le
i\le m.$
\begin{lemma}\label{c3bed}
Suppose (D1) and (D2) are satisfied. For all $x\in \R^m$, $h \not=
0$ and $i \in \{1, \ldots, m\}$, let $\zeta^{x,h,i} = \frac1h
(F(X^{x+h e_i}_T) - F(X^x_T))$. Then for every $p > 1$ there
exists a $C>0$, dependent only on $p$ and the bounds of $b,\rho,F$ and its derivatives, such that for all
$x$, $x' \in \R^m$ and $h$, $h' \not= 0$, \ben
E\Big[|\zeta^{x,h,i} - \zeta^{x',h',i}|^{2p} \Big] \le C (|x-x'|^2
+|h - h'|^2)^p. \een
\end{lemma}
\begin{proof}
Note that by Ito's formula $F(X^x_t)= F(X^x_0) + \int_0^t \nabla
F(X^x_s) \cdot \rho(s, X^x_s) dW_s + \int_0^t \cL F ds$. Thus
$F(X^x_t)$ is a diffusion with coefficients $\tilde
\rho(s,x)=\nabla F(x) \cdot \rho(s, x)$ and $\tilde
b(s,x)=\sum_{i=1}^m b_i(s,x) \frac{\partial F(x)}{\partial x_i} +
\frac12 \sum_{i, j=1}^m \rho_{ij}(s,x) \frac{\partial^2
F(x)}{\partial x_i \partial x_j}$, $(s,x)\in[0,T]\times \IR^m$. By
(D2) we have $\tilde \rho \in \bf B^{1\times d}$ and $\tilde b
\in \bf B^{1\times 1}$. Therefore, by using standard results on
stochastic flows (see Lemma 4.6.3 in \cite{kunita}), we obtain the
result.
\end{proof}
Notice that since $F$ is bounded and growth condition
(\ref{subquadra}) holds, there exists a unique  solution $(Y^x,
Z^x) \in \cS^\infty(\IR) \otimes \cH^2(\IR^d)$ of the BSDE in
(\ref{fbsde}) for all $x \in \IR^d$. One can even show that we may
choose the family $(Y^x)_{x\in\IR^m}$ such that it is continuous
in $x$.
\begin{lemma} \label{300307-2}
Let (D1), (D2), (\ref{subquadra}) and (\ref{lipabl}) be satisfied, and assume that $F$ is
bounded and that there exists a constant $K\in\R_+$ such that for all $t\in [0,T]$, $x, x' \in $ and $z\in \R^d$
\ben\label{lipinx}
|f(t,x,z) - f(t,x',z)| \le K (1+|z|) |x-x'|. 
\een
Then for all $p>1$ there exists a constant $C \in \IR_+$
such that for all $x$, $x' \in \IR^m$,
\begin{equation} \label{lipsup}
 E\sup_{t\in[0,T]}|Y^{x}_t - Y^{x'}_t|^{2p} \le C |x-x'|^{2 p},
\end{equation}
\begin{equation}
 E\Big[\left(\int_0^T |Z^{x}_t - Z^{x'}_t|^2 dt \right)^{p}\Big] \le C |x-x'|^{2p}.
\end{equation}
In particular, Kolmogorov's continuity criterion implies that
there exists a measurable process $\widetilde Y:
\Omega\times[0,T]\times\R^m$ such that $(t,x) \mapsto \widetilde
Y^x_t$ is continuous for a.a. $\om$; and for all $(t,x)\in
[0,T]\times \R^m$ we have $\widetilde Y^x_t = Y^x_t$ a.s.
\end{lemma}
\begin{proof}
For $\alpha \in \R$, let $\chi(y) = e^{\alpha y}$.
By applying Ito's formula to $\chi(Y^x)$ and using standard
arguments one can show that $\int_0^\cdot Z^x dW \in \bmo$
with the BMO norm depending only on the bound of $F$ and the
growth constant of $f$ in $z$.

For all $x, x' \in\R^m$ let $U_t = Y^{x}_t - Y^{x'}_t$, $V_t =
Z^{x}_t - Z^{x'}_t$ and $\zeta = F(X^x)-F(X^{x'})$. We use a line
integral transformation in order to show that $U^x$ can be seen as
a BSDE with generator satisfying a Lipschitz condition of the type
(\ref{bmolipschitz}). Define $J_t = \int_0^1 \nabla_x f(t,X^{x}_t - \theta(X^x_t - X^{x'}_t), Z^{x}_t) d
\theta$ and $H_t = \int_0^1 \nabla_z f
(t,X^{x'}_t, Z^{x'}_t - \theta(Z^x_t - Z^{x'}_t)) d \theta$ and
observe that
\begin{eqnarray*}
U_t &=& \zeta - \int_t^T V_s dW_s + \int_t^T (f(s,X^x_s, Z^x_s) - f(s,X^{x'}_s, Z^x_s)) + (f(s,X^{x'}_s, Z^x_s)-f(s,X^{x'}_s, Z^{x'}_s))ds \\
&=& \zeta - \int_t^T V_s dW_s + \int_t^T (J_s (X^x_s-X^{x'}_s) + H_s V_s) ds.
\end{eqnarray*}
The moment estimate of Lemma \ref{apriori-quadra} applied to the
pair $(U,V)$ leads to \be E \Big[\sup_{t\in[0,T]} |U_t|^{2p}
\Big]+ E \left[ \left(\int_0^T |V_s|^2 d s \right)^{p} \right]
\leq C \, \left( E\Big[\,|\zeta|^{2p q} +(\int_0^T |J_s(X^x_s -
X^{x'}_s)| d s)^{2pq} \Big]\right)^{\frac{1}{q}}, \ee for some
constants $C>0$ and $q>1$. 
By (\ref{lipinx}) we have $\nabla_x f(t,x,z) \le K (1+|z|)$, and hence
\be 
E(\int_0^T |J_s(X^x_s - X^{x'}_s)| d s)^{2pq} \le K^{2pq}\left(E( \int_0^T (1+|Z^x_s|)^2 d s)^{2pq}\right)^\frac12 \left(E(\int_0^T |X^x_s - X^{x'}_s|^2 d s)^{2pq}\right)^\frac12.
\ee
Lemma \ref{apriori-quadra} implies that $E( \int_0^T (1+|Z^x|)^2 d s)^{2pq}$ is bounded, and by standard results on moment estimates of SDEs we have
$E(\int_0^T |X^x_s - X^{x'}_s|^2 d s)^{2pq} \le C'|x-x'|^{4pq}$ for some constant $C'\in\R_+$ (see Theorem 3.2 in \cite{raokunita}). Moreover, the Lipschitz property of $F$ guarantees that there exists a constant $C''$ such that $E|\zeta|^{2p q} \le C''|x-x'|^{2pq}$, and hence the desired result follows.
\end{proof}

The following theorem guarantees pathwise continuous
differentiability of an appropriate modification of the solution
process.
\begin{theo}\label{diffbsde}
Let (D1), (D2), (\ref{subquadra}) and (\ref{lipabl}) be satisfied, and suppose that $F$ is
bounded and $f$ satisfies (\ref{lipinx}). Besides suppose that  $\nabla_z f$ is globally Lipschitz continuous in $(x,z)$ and that $\nabla_x f$ satisfies for all $t\in [0,T]$, $x, x' \in $ and $z, z' \in \R^d$
\ben\label{lipinnablax}
|\nabla_x f(t,x,z) - \nabla_x f(t,x',z')| \le K (1+|z| + |z'|) (|x-x'| + |z-z'|). 
\een
Then there exists a function
$\Omega \times [0, T] \times \R^m \to \R^{m+1+d}$, $(\om, t, x)
\mapsto (X^x_t, Y^{x}_t, Z^x_t)(\om)$, such that for almost all
$\om$, $X^x_t$ and $Y^x_t$ are continuous in $t$ and continuously
differentiable in $x$, and for all $x$, $(X^x_t, Y^x_t, Z^x_t)$ is
a solution of FBSDE (\ref{fbsde}). Moreover, there exists a
process $\nabla_xZ^x \in \cH^2$ such that the pair $(\nabla_x
Y^x,\nabla_x Z^x)$ solves the BSDE
\begin{equation}\label{ablbsde}
\begin{array}{lll}
\nabla_x Y_t^{x}
&=& \nabla_x  F(X_T^{x}) \nabla_x X_T^{x} - \int_t^T \nabla_x Z_s^{x} d W_s \\
&& \qquad + \int_t^T \left[ \nabla_x f(s,X_s^{x},Z_s^{x}) \nabla_x X_s^{x} + \nabla_z f(s,X_s^{x},Z_s^{x})
\nabla_x Z_s^{x} \right] d s.
\end{array}
\end{equation}
\end{theo}
We will use Kolmogorov's Lemma in order to prove the theorem. Let
$x\in \IR^m$. For all $h \not= 0$, let $\Delta^{x,h}_t =
\frac1h(X^{x+ h e_i}_t - X^x_t)$, $U^{x,h}_t = \frac1h (Y^{x+e_i
h}_t - Y^x_t)$, $V^{x,h}_t = \frac1h (Z^{x+he_i}_t - Z^x_t)$ and
$\zeta^{x,h} = \frac1h (\xi(x+h e_i) - \xi(x))$. We need the
following estimates.
\begin{lemma}\label{lemma.kolmo} For all $p>1$, $x, x'\in\R^m, h,
h'\not=0$ we have with some constant $C$
\begin{equation}\label{kolmo}
E\Big[\sup_{t\in [0,T]} |U^{x,h}_t - U^{x',h'}_t|^{2p}\Big] \le C (|x-x'|^2+|h-h'|^2)^p.
\end{equation}
\end{lemma}
\begin{proof}
Let $p > 1$. Note that for all $h\not=0$ \be U^{x,h}_t &=&
\zeta^{x,h} - \int_t^T V^{x,h}_s d W_s + \int_t^T \frac1h [f(s,X^{x+h e_i}_s,Z^{x+h e_i}_s)-f(s,X^x_s,Z^x_s)] d s. \ee We
use a line integral transformation in order to show that $U^{x,h}$
can be seen as a BSDE with random Lipschitz bound. To this end
define two $(\cF_t)$-adapted processes by \be
A^{x,h}_t &=& \int_0^1 \nabla_x f(t, X^x_t+\theta(X^{x+h e_i}_t-X^x_t),Z^{x}_t) d\theta, \\
I^{x,h}_t &=& \int_0^1 \nabla_z f(t,X^{x+h e_i}_t,
Z^x_t+\theta(Z^{x+h e_i}_t-Z^x_t)) d\theta. \ee Then
\begin{eqnarray*}
\frac1h [f(t,X^{x+h e_i}_t,Z^{x+h e_i}_t)-f(t,X^x_t,Z^x_t)] &=& A^{x,h}_t \Delta^{x,h}_t  + I^{x,h}_t
V^{x,h}_t.
\end{eqnarray*}
The growth condition (\ref{lipabl}) implies that $|I^{x,h}| \le
M(1 + |Z^x| + |Z^{x+h e_i}|)$, and hence $\int_0^\cdot I^{x,h}
dW \in \bmo$. Thus we obtain a BSDE with generator satisfying
condition (\ref{bmolipschitz}).

Now let $x, x'\in\R^m$ and $h, h'\not= 0$. Then the difference
$(U^{x,h} - U^{x',h'}, V^{x,h} - V^{x',h'})$ solves again a BSDE
with generator of the type (\ref{bmolipschitz}), namely \be y_t
&=& \zeta^{x,h} -
\zeta^{x',h'} - \int_t^T z_s dW_s\\
&&\quad - \int_t^T (I^{x,h} z_s + (I^{x,h}_s -
I^{x',h'}_s)V^{x',h'} + A^{x,h}_s \Delta^{x,h}_s- A^{x',h'}_s
\Delta^{x',h'}_s) ds. \ee Therefore Lemma \ref{apriori-quadra}
yields for $q>1$ \ben \label{261006-11para}
&&E\Big[\sup_{t\in [0,T]} |U^{x,h}_t - U^{x',h'}_t|^{2p}\Big]
\le C \Big\{ E\Big[|\zeta^{x,h} - \zeta^{x',h'}|^{2p q}\nonumber\\
&&\qquad\qquad\qquad +
E\Big[\left(\int_0^T (|A^{x,h}_s \Delta^{x,h}_s- A^{x',h'}_s
\Delta^{x',h'}_s| + |I^{x,h}_s - I^{x',h'}_s| |V^{x',h'}|) ds
\right)^{2p q}\Big]^{\frac{1}{q}} \Big\}. \nonumber \een

To treat the first term on the right hand side, use Lemma
\ref{c3bed} to see that $E[|\zeta^{x,h} - \zeta^{x',h'}|^{2p
q}]^{\frac{1}{q}} \le C (|x-x'|^2+|h-h'|^2)^p$.

For the second term, recall that $\nabla_z f$ is Lipschitz
continuous, say with Lipschitz constant $L \in \IR_+$. We
therefore have for any $t\in[0,T]$ \be |I^{x,h}_t - I^{x',h'}_t|
&\le& L(|X^{x,h}_t - X^{x',h'}_t|+|Z^{x}_t - Z^{x'}_t| + |Z^{x+h
e_i}_t - Z^{x'+ h' e_i}_t|). \ee Now Cauchy-Schwarz' inequality
leads to \be
&&E\Big[\left(\int_0^T |I^{x,h}_s - I^{x',h'}_s||V^{x',h'}| ds \right)^{2p q}\Big]^{\frac{1}{q}} \\
&\le& \left(E\Big[\left(\int_0^T |I^{x,h}_s - I^{x',h'}_s|^2 ds
\right)^{2p q}\Big]  E\Big[\left(\int_0^T |V^{x,h}|^2 ds
\right)^{2p q} \Big]\right)^{\frac{1}{2q}}. \ee So Lemma
\ref{300307-2} and Lemma 4.5.6 in \cite{kunita} imply with some
constant $C$ \be E\Big[\left(\int_0^T |I^{x,h}_s - I^{x',h'}_s|^2
ds \right)^{2p q}\Big]^{\frac{1}{2q}} &\le& C
(|x-x'|^2+|h-h'|^2)^p. \ee The term $E\Big[\left(\int_0^T
|V^{x,h}|^2 ds \right)^{2p q} \Big]$ is seen to be bounded by an
appeal to Lemma \ref{300307-2}.

It remains to show that $E\Big[\left(\int_0^T |A^{x,h}_s \Delta^{x,h}_s- A^{x',h'}_s
\Delta^{x',h'}_s| ds \right)^{2p q}\Big]^{\frac{1}{q}} \le C
(|x-x'|^2+|h-h'|^2)^p.$
For this we separately estimate the two summands on the right hand side of the following inequality 
\be
|A^{x,h}_s \Delta^{x,h}_s- A^{x',h'}_s 
\Delta^{x',h'}_s| \le |A^{x,h}_s| |\Delta^{x,h}_s - \Delta^{x',h'}_s| + |\Delta^{x',h'}_s| |A^{x,h}_s - A^{x',h'}_s|.
\ee
First note that due to (\ref{lipinx}) we have for some constants $C_1, C_2 \ldots$ 
\be
\int_0^T |A^{x,h}_s| |\Delta^{x,h}_s - \Delta^{x',h'}_s| ds \le C_1 \left( \int_0^T (1+|Z^x_s|)^2 ds \right)^\frac12
\left(\int_0^T |\Delta^{x,h}_s - \Delta^{x',h'}_s|^2 ds\right)^\frac12, 
\ee
which implies, together with Lemma \ref{apriori-quadra} and standard estimates of differences of the $\Delta^{x,h}$ (see Theorem 3.3 in \cite{raokunita}),  
\be
E\left(\int_0^T |A^{x,h}_s| |\Delta^{x,h}_s - \Delta^{x',h'}_s| ds \right)^{2pq} &\le& C_2 \left(E\left(\int_0^T |\Delta^{x,h}_s - \Delta^{x',h'}_s|^2 ds\right)^{2pq}\right)^\frac12 \\
&\le& C_3 (|x-x'|^2+|h-h'|^2)^{pq}.
\ee
Secondly, from (\ref{lipinnablax}) we obtain 
\be
& & \int_0^T |\Delta^{x',h'}_s| |A^{x,h}_s - A^{x',h'}_s| ds \\
&\le& C_4 \int_0^T (1+|Z^x_s| + |Z^{x'}_s|) \left(|X^{x}_s - X^{x'}_s| + |X^{x+h e_i}_s - X^{x'+h'e_i}_s|+|Z^{x}_s - Z^{x'}_s| \right)ds 
\ee
and hence, with Lemma \ref{300307-2} and moment estimates for $X^{x}$,
\be
& & E\left(\int_0^T |\Delta^{x',h'}_s| |A^{x,h}_s - A^{x',h'}_s| ds \right)^{2pq} \\
&\le& C_5 \left( E \left(\int_0^T \left(|X^{x}_s - X^{x'}_s| + |X^{x+h e_i}_s - X^{x'+h'e_i}_s|+|Z^{x}_s - Z^{x'}_s| \right)^2 ds \right)^{2pq}\right)^\frac12 \\
&\le& C_6(|x-x'|^2+|h-h'|^2)^{pq}.
\ee 
Combining the estimates just derived, we conclude
\[
E\Big[\sup_{t\in [0,T]} |U^{x,h}_t - U^{x',h'}_t|^{2p}\Big] \le C
(|x-x'|^2+|h-h'|^2)^p.
\]
This completes the proof of the lemma.
\end{proof}
\begin{proof}[Proof of Theorem \ref{diffbsde}]
Note that by Lemma \ref{300307-2} we may assume that $(t,x)
\mapsto Y^x_t$ is continuous for all $\om$. Then $U^{x,h}$ has
continuous paths for all $x\in \R^m$ and $h \not= 0$.

Let $\cQ$ be the collection of all pairs $(x,h)$ where $x$ is a
vector of dyadic rationals in $\R^m$ and $h \not= 0$ a dyadic
rational in $\R$. Since inequality (\ref{kolmo}) is valid,
Kolmogorov's lemma implies that there exists a null set $N$ such
that for all $\om \in N^c$ the function $\cQ \ni (x,h) \mapsto
U^{x,h}$ can be uniquely extended to a continuous function from
$\R^{m+1}$ into the space of continuous functions endowed with the
sup norm (see Thm 73, Ch. IV, \cite{protter}). Such a null set $N$
can be chosen for any direction $i$ in which we differentiate, and
hence there exists a modification of $Y^x$ such that for all $t$
the mapping $x \mapsto Y^x_t$ possesses continuous partial derivatives.

Finally it is straightforward to show that the derivative $\nabla_x Y^x$ together with a process $\nabla_x Z^x$,
defined as an $\cH^2$ limit of the processes $V^{x,h}$ as $h \to 0$, solve the BSDE (\ref{ablbsde}).
\end{proof}
\subsection{The Markov property of FBSDE}
The forward part of our FBSDE (\ref{fbsde}) is solved by a time
inhomogeneous Markov process. We will now investigate the
consequences of this fact in more detail. Let us fix an initial
time $t\in[0,T),$ as well as an initial state $x$ to be taken by
our forward process at this time. Then, conditioned on taking the
value $x$ at time $t$, the forward process satisfies the SDE \ben
\label{fsde} X^{t,x}_s = x+\int_t^s b(r,X^{t,x}_r) d r+\int_t^s
\rho(r,X^{t,x}_r) d W_r, \een where $x\in\R^m$ and $s \in [t,T]$.
We will assume that the coefficients satisfy a growth and a
Lipschitz condition. More precisely, assume that there exists a
constant $C\in \R_+$ such that for all $x$, $x' \in \R^m$ and
$t\in[0,T]$
\begin{equation} \label{lipbesch}
 \begin{array}{ccl} |b(t,x) - b(t,x')| + |\rho(t,x) - \rho(t,x')| & \le & C(|x-x'|), \\
  |b(t,x)| + |\rho(t,x)| & \le & C(1 +|x|). \end{array}
\end{equation}
Condition (\ref{lipbesch}) guarantees that there exists a unique
solution of (\ref{fsde}). It moreover implies that $X^{t,x}_r$ is
Malliavin differentiable and that its Malliavin gradient has a
representation involving, for $(t,x)$ fixed, the global flow on
the space of nonsingular linear operators $\Phi^{t,x}$ on $\R^m$
defined by the equation
$$\Phi^{t,x}_s = 1_{\R^m} + \int_t^s \nabla_x b(u, X_u^{t,x}) \Phi_u^{t,x}
du + \int_t^s \nabla_x \rho(u, X_u^{t,x}) \Phi_u^{t,x} d W_u,\quad
s\ge t.$$ Here $\nabla_x b$ and $\nabla_x \rho$ describe the
gradients of $b$ resp. $\rho$ existing in the weak sense under
(\ref{lipbesch}), $1_{\R^m}$ the $m\times m$ unit matrix. The
Malliavin gradient is then given by the formula (see Nualart
\cite{nualart}, p. 126) \begin{equation}\label{mallgradient}
 D_{\theta} X_s^{t,x} =
\Phi_s^{t,x} (\Phi_{\theta}^{t,x})^{-1}\, \rho (\theta,
X_{\theta}^{t,x}),\quad t\le \theta\le s.\end{equation}

With the Markov process $X^{t,x}$ starting at time $t$ in $x$ in
mind, we now consider BSDE of the form
\begin{equation} \label{bsde}
Y^{t,x}_s = F(X^{t,x}_T) - \int_s^T Z^{t,x}_r dW_r + \int_s^T f(r,X^{t,x}_r, Z^{t,x}_r) dr.
\end{equation}
In accordance with Section \ref{controlviabsdes}, we now assume
that the generator is a {\em deterministic} Borel measurable
function $f: [0,T] \times \R^m \times \R^{d} \to \R$. Again we
assume that $f$ is differentiable in $(x,z)$ and that there exists
a constant $M \in \IR_+$ such that for all $(t,x,z) \in
[0,T]\times\R^m\times \R^d$ we have
\begin{equation} \label{ex+eind}
|f(t,x,z)| \le M(1 + |z|^2)  \,\,\,\, \textrm{
a.s.}\qquad\textrm{and}\qquad |\nabla_z f(t,x,z)|  \le M(1 +
|z|)\,\,\,\,\textrm{ a.s.}
\end{equation}
for all $(t,z) \in [0,T] \times \IR^m$. If $F$ is bounded, then it
follows from Theorem 2.3 and 2.6 in \cite{koby} that there exists
a unique solution $(Y^{t,x}, Z^{t,x}) \in \cS^\infty(\R) \otimes
\cH^2(\R^d)$ of the BSDE (\ref{bsde}). The next result states that
the solution of the BSDE is already determined by the forward
process $X^{t,x}$. In order to formulate it, for all $m \in \IN$
we denote by $\cD^m$ the $\sigma$-algebra on $\IR^m$ generated by
the family of functions $\IR^m \ni x \mapsto E\int_t^T \varphi(s,
X^{t,x}_s) ds$, where $t\in[0,T]$ and $\varphi: [0,T]\times \IR^m
\to \IR$ is bounded and continuous.
\begin{theo} \label{markov}
Let $F: \R^m \to \R$ be a bounded Borel function, suppose that $f$
satisfies (\ref{ex+eind}) and the coefficients of the forward
diffusion (\ref{lipbesch}). Suppose that there exist functions
$f_n:[0,T] \times \R^m \times \IR^d \to \IR$, globally Lipschitz
continuous in $(x,z)$, such that for almost all $\om$ and for all compact sets $K
\subset \R^m \times \IR^d$ the sequence $f_n$ converges to $f$
uniformly on $[0,T] \times K$. Then there exist two
$\cB[0,T]\otimes\cD^m$- and $\cB[0,T]\otimes \cD^m$-measurable
deterministic functions $u$ and $v$ on $[0,T]\times \R^m$ such
that
\begin{equation}
Y^{t,x}_s = u(s, X^{t,x}_s) \quad \textrm{ and } \quad Z^{t,x}_s = v(s,X^{t,x}_s) \rho(s,X^{t,x}_s),
\end{equation}
for $P\otimes \lambda$-a.a.\ $(\om, s) \in \Omega \times [t,T]$.
\end{theo}
\begin{proof}
Let $f^n$ be Lipschitz continuous in $(x,z)$ such that $f^n$
converges locally uniformly on $\IR_+ \times \R^m \times \IR^d$.
Let $(t,x)\in[0,T]\times\R^m$ and denote by $(Y^n,Z^n) =
((Y^n)^{t,x},(Z^n)^{t,x})$ the solution of the BSDE with generator
$f^n$ and terminal condition $\xi = F(X^{t,x}_T)$. It follows from
Theorem 2.8 in \cite{koby} that $Y^n$ converges to $Y^{t,x}$ in
$\cH^\infty(\R)$, and $Z^n$ converges to $Z^{t,x}$ in
$\cH^2(\R^d)$. By taking a subsequence if necessary, we may assume that $Z^n$ converges to $Z^{t,x}$ a.s. on $\Omega \times [0,T]$.

According to Theorem 4.1 in \cite{EPQ}, there exist
$\cB[0,T]\otimes\cD^m$- and $\cB[0,T]\otimes \cD^m$-measurable
deterministic functions $u_n(t,x)$ and $v_n(t,x)$ that satisfy the
representations $Y^n_s = u_n(s, X^{t,x}_s)$ and $Z^n_s =
v_n(s,X^{t,x}_s) \rho(s,X^{t,x}_s)$ for all $s\in[t,T]$ a.s. Now
define
\[ u(t,x) = \liminf_n u_n(t,x) \quad \textrm{ and } \quad v(t,x) = \liminf_n v_n(t,x). \]
Then $Y^{t,x}_s = u(s,X^{t,x}_s)$ and $Z^{t,x}_s = v(s,X^{t,x}_s) \rho(s,X^{t,x}_s)$, a.s.
\end{proof}
By combining Theorem \ref{markov} with Theorem \ref{diffbsde} we
obtain a representation of the control process $Z^{t,x}$ in terms
of the derivative of $Y^{t,x}$ with respect to $x$.
\begin{theo}\label{markov2}
Suppose that the assumptions of Theorem \ref{markov} are
satisfied. Besides assume that $\nabla_z f$ is
globally Lipschitz continuous, that (\ref{lipinx}) and (\ref{lipinnablax}) are satisfied, and further that the forward
coefficients satisfy the stronger conditions (D1) and (D2). Then
$u(t,x)$ is differentiable in $x$ for a.a. $t\in [0,T]$. Moreover,
\begin{equation} \label{zalsabl}
Z^{t,x}_s = \nabla_x u(t,X^{t,x}_s) \rho(s,X^{t,x}_s),
\end{equation}
for $P\otimes \lambda$-a.a.\ $(\om, s) \in \Omega \times [t,T]$.
\end{theo}
\begin{proof}
Recall that $X^{t,x}_s$ is Malliavin differentiable and that the
assumptions of Lemma \ref{300307-2} are satisfied. Equation
(\ref{lipsup}) implies that $x \mapsto u(t,x) = Y^{t,x}_t$ is
Lipschitz continuous. Therefore $Y^{t,x}_s = u(s, X^{t,x}_s)$ is
Malliavin differentiable (see Proposition 1.2.2 \cite{nualart}).
By Theorem \ref{diffbsde}, $u(t,x)$ is differentiable in $x$, and
by the chain rule we have $D_\theta Y_s^{t,x} = \nabla_x u(s,
X^{t,x}_s) D_\theta X^{t,x}_s$. Since due to (\ref{mallgradient})
$D_s X^{t,x}_s = \rho(s,X^{t,x}_s)$ and $Z^{t,x}_s = D_s
Y_s^{t,x}$ (the later following f.ex.\ from Lemma 5.1 in
\cite{EPQ}), Theorem \ref{markov} implies (\ref{zalsabl}).
\end{proof}

\subsection{Differentiability of Quadratic BSDE with parameterized terminal condition}

For this subsection we pass to a more abstract parameter
dependence of the solution of a BSDE than studied above in a pair
of forward and backward SDE. We consider the BSDE
\ben\label{bsde.xix} Y^x_t&=& \xi(x)-\int_t^T Z^x_s dW_s+\int_t^T
f(s,Z^x_s)ds,\qquad t\in[0,T], x\in\R^m. \een Throughout we assume
that
\begin{enumerate}[({E}1)]
\item $\R^m\ni x\mapsto\xi(x)\in \R$ is a bounded random field
which as a function of $x$ is differentiable with bounded partial
derivatives; $\nabla \xi(x)$ is also Lipschitz in $x$; also
$f(t,0)$ is $(\cF_t)-$adapted and satisfies $f(t,0)\in L^p$ for all
$p\geq 1$.

\item there exists $M\in\IR_+$ such that $|f(t,z)|\leq M(1+|z|^2)$
a.s.; $f$ is differentiable in $z$ such that $|\nabla_z
f(t,z)|\leq M(1+|z|)$ for all $(t,z)\in [0,T]\times \IR^d$ a.s.

\item for all $x\in \R^m$, $h \not= 0$ and $i \in \{1, \ldots,
m\}$, let $\zeta^{x,h,i} = \frac1h (\xi(x+he_i) - \xi(x))$. Then
for every $p > 1$ there exists a $C>0$, dependent only on $p$,
such that for all $x$, $x' \in \R^m$ and $h$, $h' \not= 0$, \ben \label{030707}
E\Big[|\zeta^{x,h,i} - \zeta^{x',h',i}|^{2p} \Big] \le C (|x-x'|^2
+|h - h'|^2)^p.\een

\end{enumerate}

Although the terminal condition does not depend on a forward
diffusion (see Lemma \ref{c3bed} for a derivation of (\ref{030707}) in a FBSDE setting), Hypothesis (E1)-(E3) allow to apply
the methods we used in Subsection \ref{subsec:diff.fbsde} and
obtain the following theorem.

\begin{theo}\label{diffbsde.xi}
Let (E1), (E2) and (E3) be satisfied. Then there exists a function
$\Omega \times [0, T] \times \R^m \to \R^{1+d}$, $(\om, t, x)
\mapsto (Y^{x}_t, Z^x_t)(\om)$, such that for almost all $\om$ ,
the process  $Y^x_t$ is  continuous in $t$ and continuously
differentiable in $x$, and for all $x$, $(Y^x_t, Z^x_t)$ is a
solution of BSDE (\ref{bsde.xix}). Moreover, there exists a
process $\nabla_x Z^x \in \cH^2(\R^{m\times d})$ such that the pair
$(\nabla_x Y^x,\nabla_x Z^x)$ solves the BSDE
\begin{equation*}
\nabla_x Y_t^{x}
= \nabla_x  \xi(x) - \int_t^T \nabla_x Z_s^{x} d W_s + \int_t^T \left[  \nabla_z f(s,Z_s^{x})
\nabla_x Z_s^{x} \right] d s.
\end{equation*}

\end{theo}
\begin{proof}
Conditions (E1) and (E3) guarantee that the solutions of the BSDE
(\ref{bsde.xix}) exist and $(Y^x,Z^x)\in \cS^\infty(\R)\otimes
\cH^2(\R^{d})$.

Condition (E1), (E2), (E3) and the BMO property of the martingale
$\int_0^\cdot Z^x dW$ allow us to prove moment estimates that
correspond to Lemma \ref{apriori-quadra}, Lemma \ref{300307-2} and
Lemma \ref{lemma.kolmo}. Hence a simple adaptation of the proof of
Theorem \ref{diffbsde} provides the proof of Theorem
\ref{diffbsde.xi}.
\end{proof}


\end{document}